\documentclass[aps,prb,twocolumn]{revtex4}
\usepackage{amsfonts}
\usepackage{bm}
\usepackage[latin9]{inputenc}
\usepackage{color}
\usepackage{verbatim}
\usepackage{amsmath}
\usepackage{amssymb}
\usepackage{graphicx}
\usepackage{esint}

\begin{document}

\title{Hidden symmetry and protection of Dirac points  on the honeycomb lattice}
\author{Jing-Min Hou$^1$\footnote{e-mail: jmhou@seu.edu.cn} and Wei Chen$^2$}
\affiliation{$^1$Department of Physics,
Southeast University, Nanjing  211189, China\\
$^2$College of Science, Nanjing University of Aeronautics and
Astronautics, Nanjing 210016, China}

\begin{abstract}
The honeycomb lattice possesses a novel energy band structure, which
is characterized by two distinct Dirac points in the Brillouin zone,
dominating most of the physical properties of the honeycomb
structure materials. However, up till now, the origin of the Dirac
points is  unclear yet. Here, we discover a hidden symmetry on the
honeycomb lattice and prove that the existence of Dirac points is
exactly protected by such hidden symmetry. Furthermore, the moving
and merging of the Dirac points and a quantum phase transition,
which have been theoretically predicted and experimentally observed
on the honeycomb lattice, can also be perfectly explained by the
parameter dependent evolution of the hidden symmetry.
\end{abstract}

\maketitle

Graphene, a honeycomb lattice of carbon atoms, has attracted
extraordinary attention in the last decade, due to its remarkable
properties and potential
applications\cite{Novoselov1,Novoselov2,Zhang, Castro}. The band
structure  of this exotic material is characterized by two distinct
Dirac points in the Brillouin zone, dominating most of its physical
results. Although there are a multitude of researches on graphene
and other honeycomb
lattices\cite{Wallace,Park0,Li,Park,Gibertini,Asano,Gomes,Gail,Bellec},
the origin of the Dirac points is still unclear. According to the
von Neumann-Wigner theorem\cite{vNW,Landau}, there must be some
symmetry to protect the Dirac points on the honeycomb lattice, while
the robustness of Dirac points during the deformation of the lattice
structure\cite{Tarruell} excludes the possibility of a point group
protection. As a result, a novel symmetry is expected to be
responsible for the Dirac points.

In this work, we unveil the mysterious story behind the Dirac points
by showing that they are exactly protected by a kind of hidden
symmetry on the lattice structure. As its name suggests, the hidden
symmetry is not so obvious as usual symmetries such as the point
group symmetry. In general it can be described by a composite
anti-unitary operator, which consists of a translation, a complex
conjugation, and a sublattice exchange, and sometimes also a local
gauge transformation and a rotation, or is the extension of the
composite operator by a mapping method. This kind of symmetry is
seldom studied before and was firstly discovered   by one of the
authors in a toy model\cite{Hou1,Hou2}.  We find that the hidden
symmetry on the honeycomb lattice evolves along with the variation
of the parameters, which can perfectly explain the moving and
merging of the Dirac points and the quantum phase transition on the
honeycomb lattice that have been theoretically
predicted\cite{Hasegawa,Zhu,Wunsch,Pereira,Montambaux2} and
experimentally observed\cite{Tarruell}.

\begin{figure*}[t]
\includegraphics[width=0.67\columnwidth]{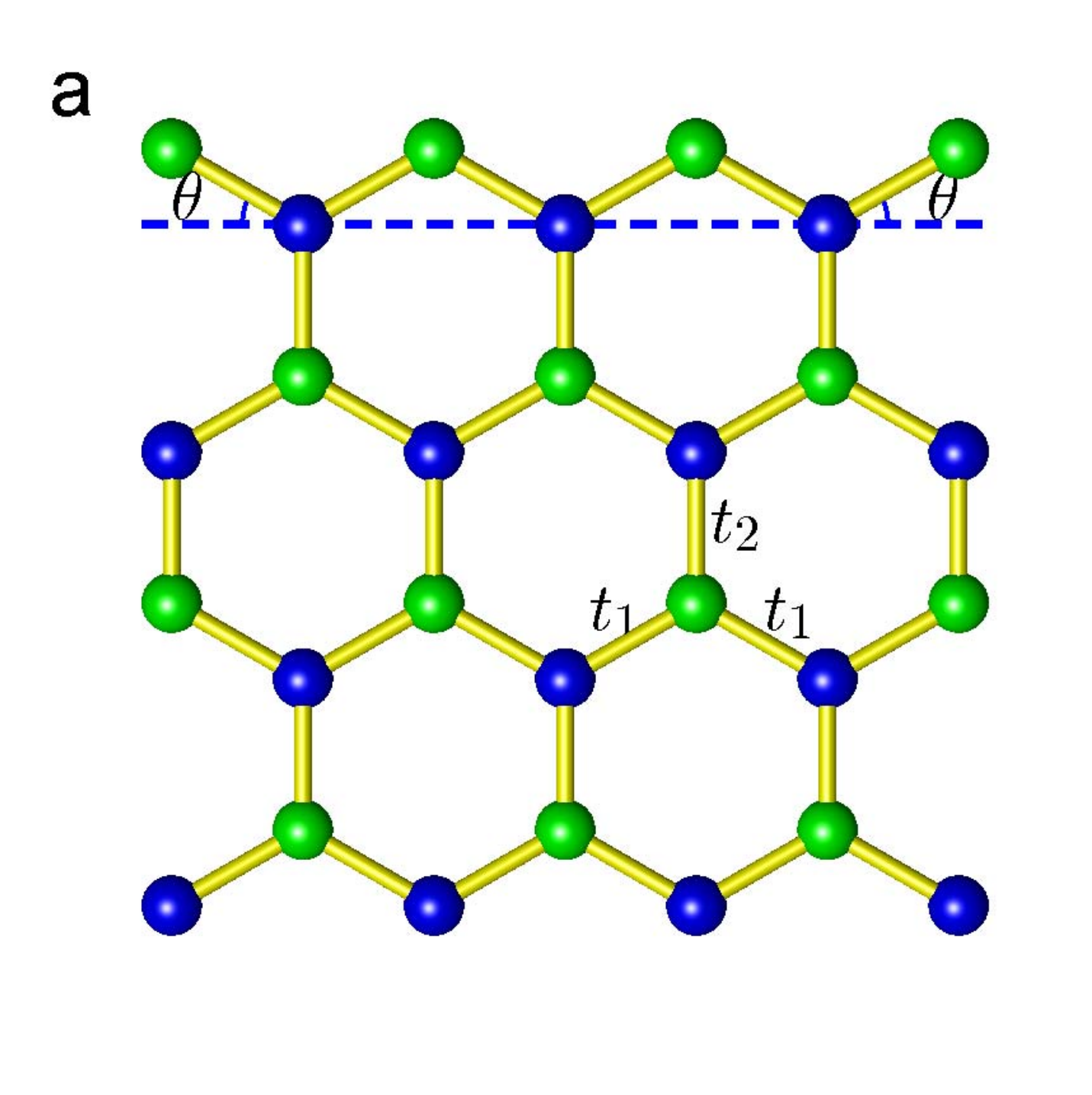}
\includegraphics[width=0.64\columnwidth]{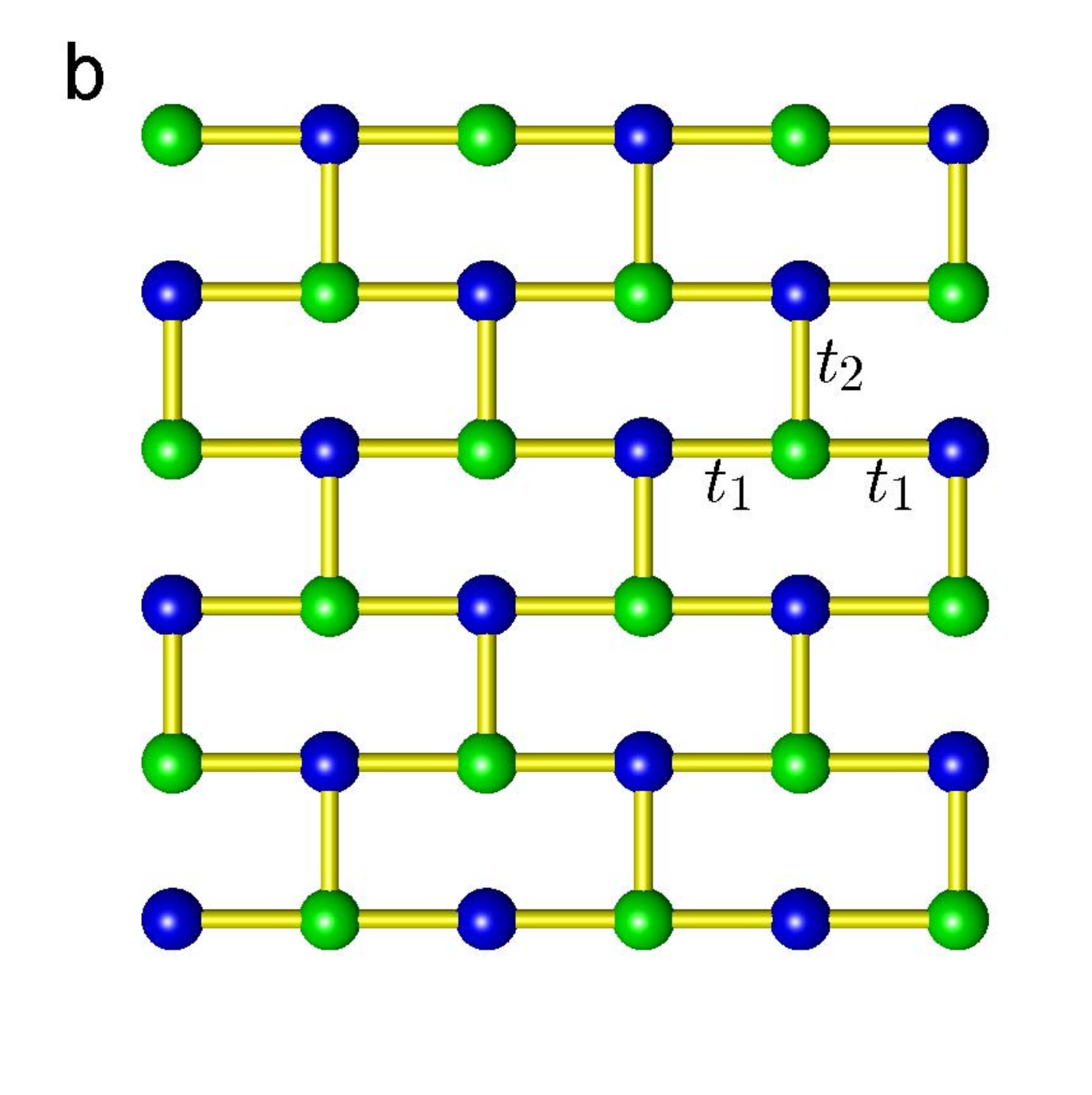}
\includegraphics[width=0.64\columnwidth]{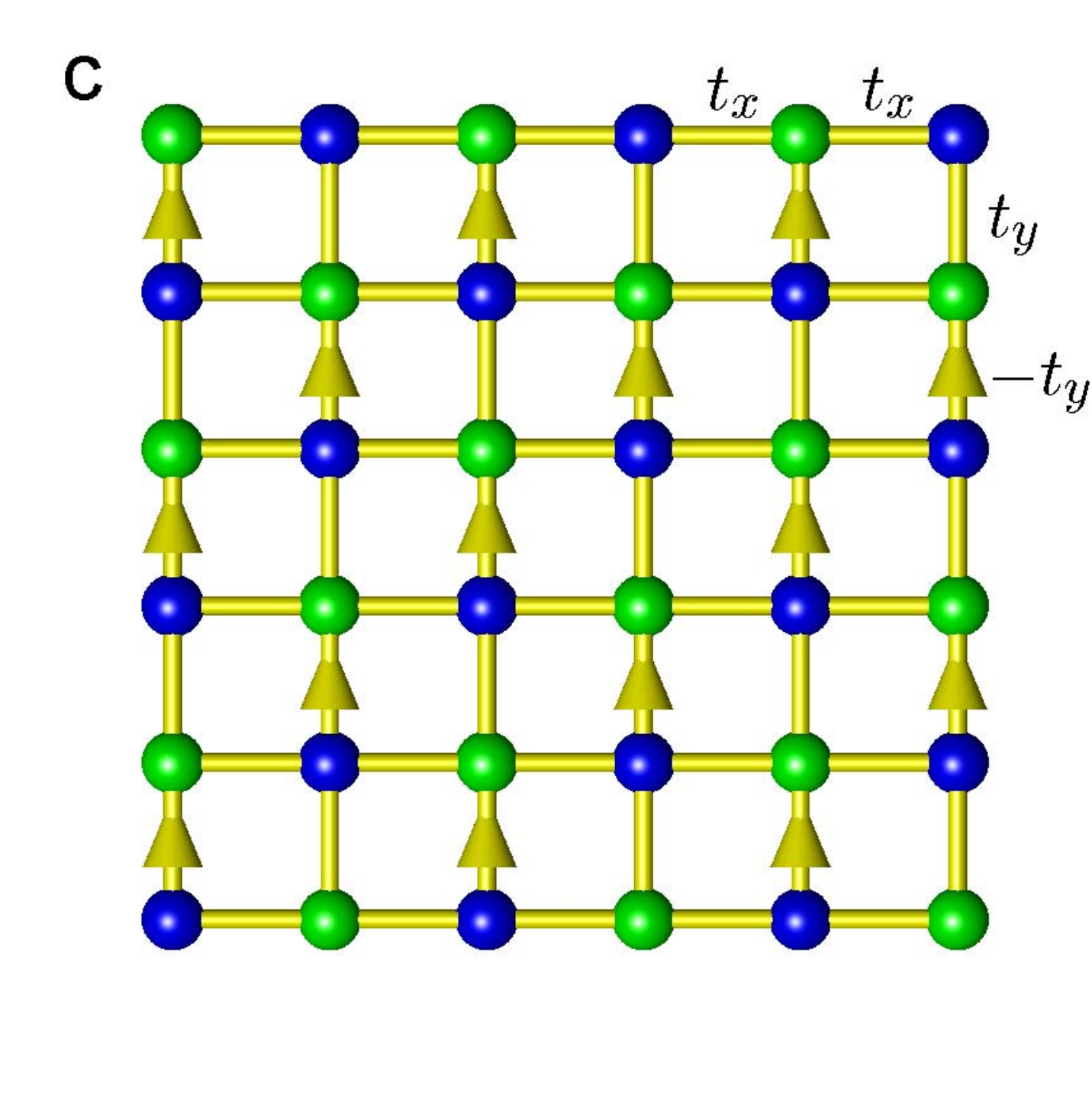}
\includegraphics[width=0.64\columnwidth]{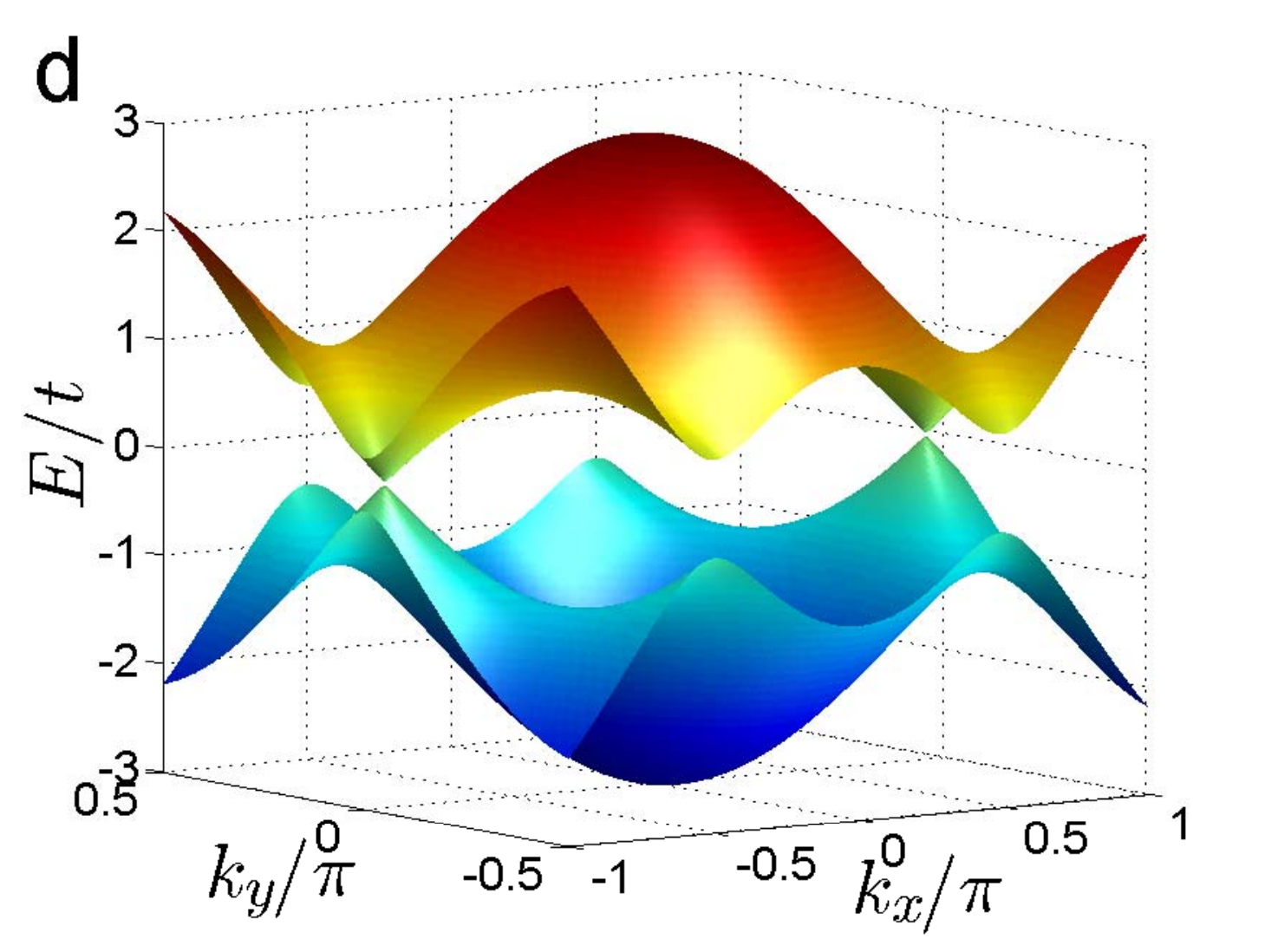}
\includegraphics[width=0.64\columnwidth]{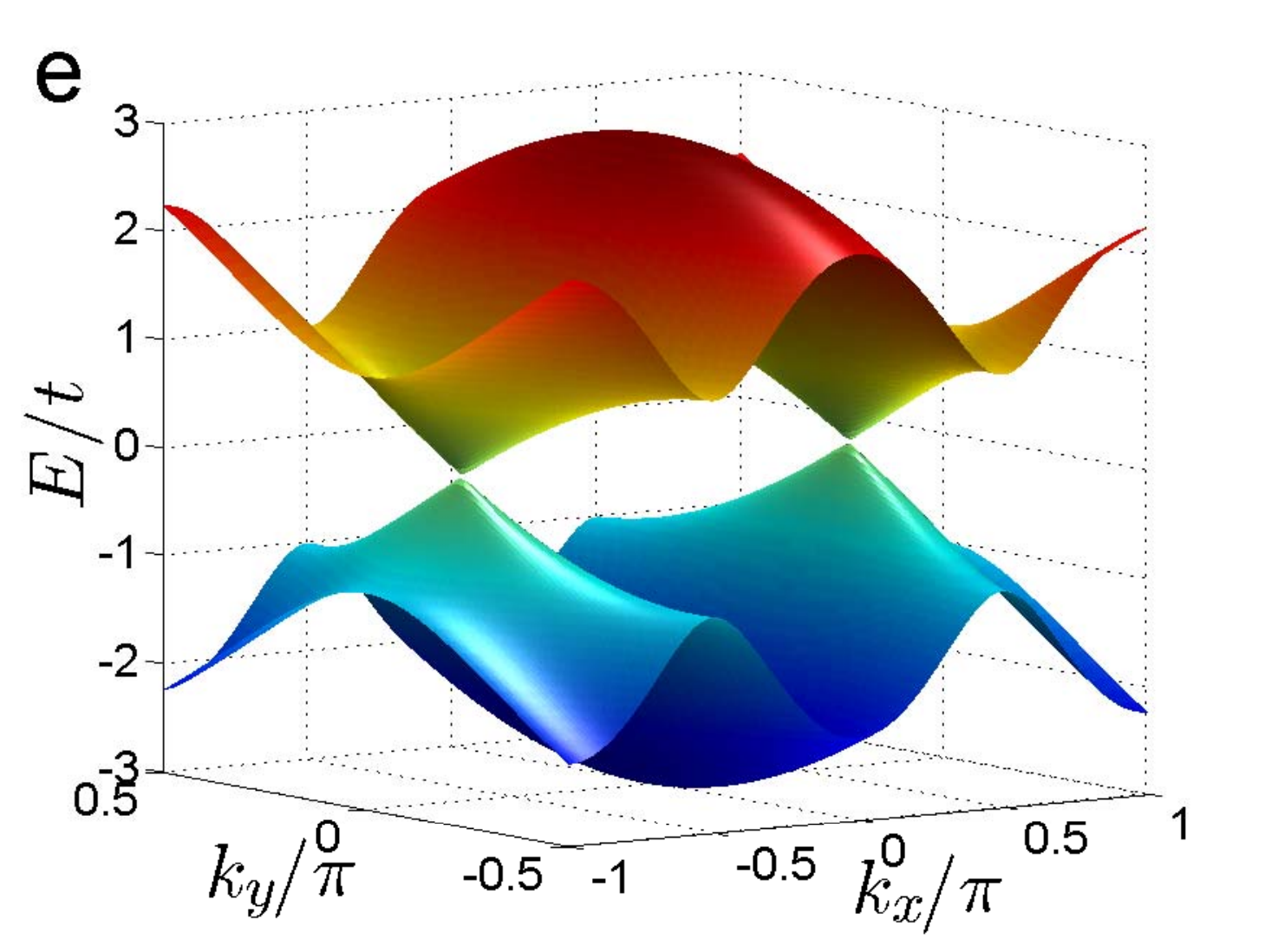}
\includegraphics[width=0.64\columnwidth]{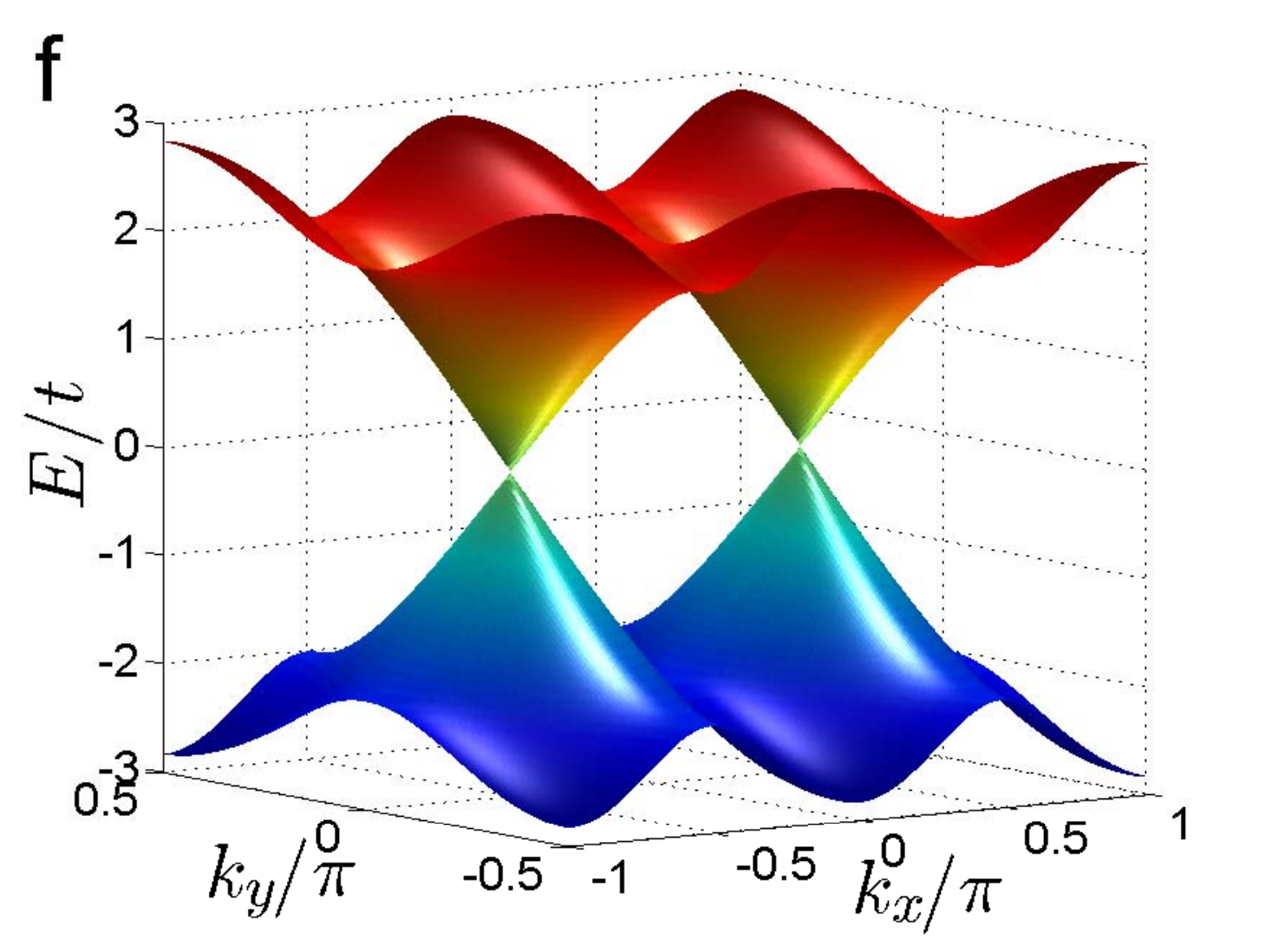}
\caption{{\bf The lattices and the dispersion relations.}
\textbf{a}, Schematic of the honeycomb lattice. $\theta$ denotes the
angle between the bonds on the zigzag line and the horizontal
direction; $t_1$ and $t_2$ represent the amplitudes of hopping.
\textbf{b}, Schematic of the brick-wall lattice, which can be
regarded as a special case of the honeycomb lattice with $\theta=0$.
\textbf{c}, Schematic of the square lattice. The arrows represent a
hopping-accompanying $\pi$ phase; $t_x$ and $t_y$ represent the
amplitudes of hopping. In \textbf{a},\textbf{b}, and \textbf{c}, the
blue and green balls represent the lattice sites in sublattices A
and B, respectively. \textbf{d}, The dispersion relation for the
honeycomb lattice model with $\theta=\pi/6$ and $t_1=t_2=t$.
\textbf{e}, The dispersion relation for the brick-wall lattice model
with $t_1=t_2=t$. \textbf{f}, The dispersion relation for the square
lattice model with $t_x=t_y=t$. } \label{fig1}
\end{figure*}

\vspace{0.5cm} \noindent \textbf{Results} \\
\noindent
\textbf{Model}. To be specific, we consider  the general honeycomb
lattice  as shown in Fig.\ref{fig1}a, where we define a bond angle
 $\theta$ as the angle between the bonds on the zigzag line and the
horizontal direction. When the bond angle takes peculiar values of
$\theta=\pi/6$ and $\theta=0$, the lattice reduces to the ideal
honeycomb lattice, such as graphene, and to the brick-wall lattice
as shown in Fig.\ref{fig1}b. The general honeycomb lattice  model
can be well described by the Bloch Hamiltonian  as (the unit bond
length is adopted)
\begin{eqnarray}
 {\cal H}_h(\mathbf{k})&=&-[ 2t_1\cos(\cos\theta k_x)\cos(\sin\theta
k_y) +t_2\cos k_y]\sigma_x\nonumber\\
& +&[2t_1\cos(\cos\theta k_x)\sin(\sin\theta k_y)-t_2\sin
k_y]\sigma_y\label{BHh}
\end{eqnarray}
where $t_1$ and $t_2$ denote the amplitudes of   hopping as sketched
in Fig.\ref{fig1}a; $\sigma_x$ and $\sigma_y$ are the pauli matrices
defined in the sublattice space.

In order to find   the hidden symmetry behind the the honeycomb
lattice, an auxiliary square lattice with a hopping-accompanying
  $\pi$ phase is introduced as well, as shown in
Fig.\ref{fig1}c, which has an intrinsic relation with   the
honeycomb lattice. The Bloch Hamiltonian for the square lattice  can
be written as
\begin{eqnarray}
{\cal H}_s(\mathbf{k})=- 2t_x\cos k_x \sigma_x -2t_y\sin
k_y\sigma_y\label{BHs}
\end{eqnarray}
 where $t_x$ and $t_y$ represent the amplitudes of hopping along the
 horizontal and vertical directions, respectively.
   (For the derivations of the Bloch
Hamiltonians of the honeycomb and square lattices see Supplementary
Section S-1.)

Geometrically, the square lattice can transform into the ideal
honeycomb lattice continuously in two steps. First, the square
lattice  changes into the brick-wall lattice when the amplitude of
hopping with a $\pi$ phase is tuned to zero, and then reaches the
ideal honeycomb lattice by a deformation of the bond angle $\theta$
from 0 to $\pi/6$, which can be understood with the help of
Fig.\ref{fig1}a-c.
  Besides the intuitive relation between these lattice structures in the real space,
their band structures also strongly correlate to each other. The
energy bands are all characterized by two linear Dirac cones in the
Brillouin zone as shown in Fig.\ref{fig1}d-f (for the definitions of
the Brillouin zones  see Supplementary Section S-2). More
importantly, these Dirac points are able to evolve continuously into
each other with the variation of the lattice parameters.  For the
general honeycomb lattice with $\beta=1$ ($\beta$ is defined as the
hopping amplitude ratio $\beta=t_2/t_1$), the corresponding Dirac
points locate  at $(\pm2\pi/3\cos\theta, 0)$ in the Brillouin zone.
As a result, when the bond angle $\theta$ varies from $\pi/6$ to
$0$, the lattice first changes from the ideal honeycomb lattice into
the brick-wall lattice, inducing a shift of the Dirac points from
$(\pm4\pi/3\sqrt{3}, 0)$ to $(\pm2\pi/3, 0)$, as shown in
Fig.\ref{fig1}d and Fig.\ref{fig1}e.  Starting from the brick-wall
lattice, the square lattice can be obtained by turning on the
amplitude of hopping with a $\pi$ phase from $0$ to $-t_y$.
Accordingly,  the corresponding Dirac points evolve from
$(\pm2\pi/3, 0)$ to $(\pm\pi/2, 0)$, as shown in Fig.\ref{fig1}f. It
is impressive that during the whole evolution of the lattice, the
Dirac points are always stable without any gap opening. We will show
that this property can be perfectly explained by the protection of
the hidden symmetry of the lattice structures.

\vspace{0.5cm}

\noindent \textbf{Hidden symmetry  and protection of Dirac points}.
Firstly, we consider the auxiliary square lattice as shown in
Fig.\ref{fig1}c. One
  can verify that the square lattice is invariant
under the action of the operator defined as
\begin{eqnarray}
 \Upsilon=(e^{i\pi})^{i_y}\sigma_x K T_{\hat{x}}
 \end{eqnarray}
  where
$T_{\hat{x}}$ is a translation operator that moves the lattice along
the horizontal direction by a unit vector $\hat{x}$; $K$  is the
complex conjugate operator; $\sigma_x$ is the Pauli matrix
representing the sublattice exchange; $(e^{i\pi})^{i_y}$ is a local
$U(1)$ gauge transformation (for details see Supplementary Section
S-3). This kind of transformation invariance  indicates a hidden
symmetry of the square lattice\cite{Hou1}. It is easy to prove that
the symmetry operator $\Upsilon$ is anti-unitary, and its square is
equal to $\Upsilon^2=T_{2\hat{x}}$.

Mathematically, the hidden symmetry operator $\Upsilon$ can be
considered as a self-mapping of the square lattice model defined as
\begin{eqnarray*}
\Upsilon:(\mathbf{k}, {\cal H}_s(\mathbf{k}),
\Psi_{s,\mathbf{k}}(\mathbf{r}))\mapsto (\mathbf{k}', {\cal
H}_s(\mathbf{k}'), \Psi_{s,\mathbf{k}'}'(\mathbf{r}))
\end{eqnarray*}
where $\Psi_{s,\mathbf{k}}(\mathbf{r})$ and
$\Psi'_{s,\mathbf{k}'}(\mathbf{r})$ are the Bloch functions of the
square lattice model.
 Performing the hidden symmetry transformation on the Bloch function
 leads to
$\Upsilon\Psi_{s,\mathbf{k}}(\mathbf{r})={\Psi'_{s,\mathbf{k}'}}(\mathbf{r})$
with $
 \mathbf{k}'=(k_x',k_y')=(-k_x, -k_y+\pi)
 $ (Methods).
If $\mathbf{k}'=\mathbf{k}+\mathbf{K}^s_m$, where $\mathbf{K}^s_m$
is a reciprocal lattice vector for the square lattice, then we can
say that $\mathbf{k}$ is a  $\Upsilon$-invariant point. In the
Brillouin zone, the $\Upsilon$-invariant points are
$\mathbf{M}_{1,2}=( \pm\pi/2, 0)$  and
$\mathbf{M}_{3,4}=(0,\pm\pi/2)$. After the hidden symmetry operator
acts on the Bloch function twice, we have
  $\Upsilon^2\Psi_{s,\mathbf{k}}(\mathbf{r})=T_{2\hat{x}}\Psi_{s,\mathbf{k}}(\mathbf{r})=e^{-2ik_x}\Psi_{s,\mathbf{k}}(\mathbf{r})$.
  Therefore, the square of the hidden symmetry operator takes a value of $\Upsilon^2=e^{-2ik_x}$ in the Bloch representation.
  For the $\Upsilon$-invariant points,  we have $\Upsilon^2=-1$ at $\mathbf{M}_{1,2}$,
while $\Upsilon^2=1$ at $\mathbf{M}_{3,4}$.  Since $\Upsilon$ is an
anti-unitary operator,
  we arrive at the important conclusion that the band energy must be
degenerate at the $\Upsilon$-invariant points $\mathbf{M}_{1,2}$,
which are just the locations of the Dirac points as shown in
Fig.\ref{fig1}f. From the above discussion, one can see that the
Dirac points on the auxiliary square lattice are exactly protected
by the
 hidden symmetry $\Upsilon$. (For details see Supplementary Section
S-5A).

  In the following, we
show that the hidden symmetry of the honeycomb lattice can be
derived from that of the auxiliary square lattice. We define a
mapping from the honeycomb lattice model to the square lattice model
as
\begin{eqnarray*}
\Omega_{\theta,\beta}:(\mathbf{k}, {\cal H}_h(\mathbf{k}),
\Psi_{h,\mathbf{k}}(\mathbf{r}))\mapsto (\mathbf{K}, {\cal
H}_s(\mathbf{K}), \Psi_{s,\mathbf{K}}(\mathbf{r}))
\end{eqnarray*}
which depends on the bond angle $\theta$ and the hopping amplitude
ratio $\beta$, with $\Psi_{h,\mathbf{k}}(\mathbf{r})$ being the
Bloch functions of the honeycomb lattice model. The operator
$\Omega_{\theta,\beta}$ maps the Bloch Hamiltonian of the honeycomb
lattice into that of the square lattice, which can be expressed as
\begin{eqnarray}
\Omega_{\theta,\beta}{\cal
H}_h(\mathbf{k})\Omega_{\theta,\beta}^{-1}={\cal H}_s(\mathbf{K})
\end{eqnarray}
where the hopping amplitudes are related by $t_y=t_2/2$ and
$t_x=t_1+t_2/2$, and the wave vector $\mathbf{K}=(K_x,K_y)$ is
defined by $\mathbf{k}$ as
\begin{eqnarray*}
K_x&=&
\begin{cases}
-\mathcal{K}_{\theta,\beta}(\mathbf{k}),&\mbox{for}\  k_x\leq0\\
\mathcal{K}_{\theta,\beta}(\mathbf{k}),& \mbox{for} \ k_x\geq0
\end{cases}\\
K_y&=&(1+\sin\theta)k_y
\end{eqnarray*}
with
$\mathcal{K}_{\theta,\beta}(\mathbf{k})\equiv\arccos\{\frac{2}{2+\beta}\cos
(\cos\theta k_x)+\frac{ \beta }{2+\beta}\cos [(1+\sin\theta)k_y]\}$.
 Performing the mapping procedure on the Bloch function
as well, one obtains
$\Omega_{\theta,\beta}\Psi_{h,\mathbf{k}}(\mathbf{r})=\Psi_{s,\mathbf{K}}(\mathbf{r})$.
(Interpreting this mapping in an intuitive way see Supplementary
Section S-4).

\begin{figure}[t]
\includegraphics[width=0.95\columnwidth]{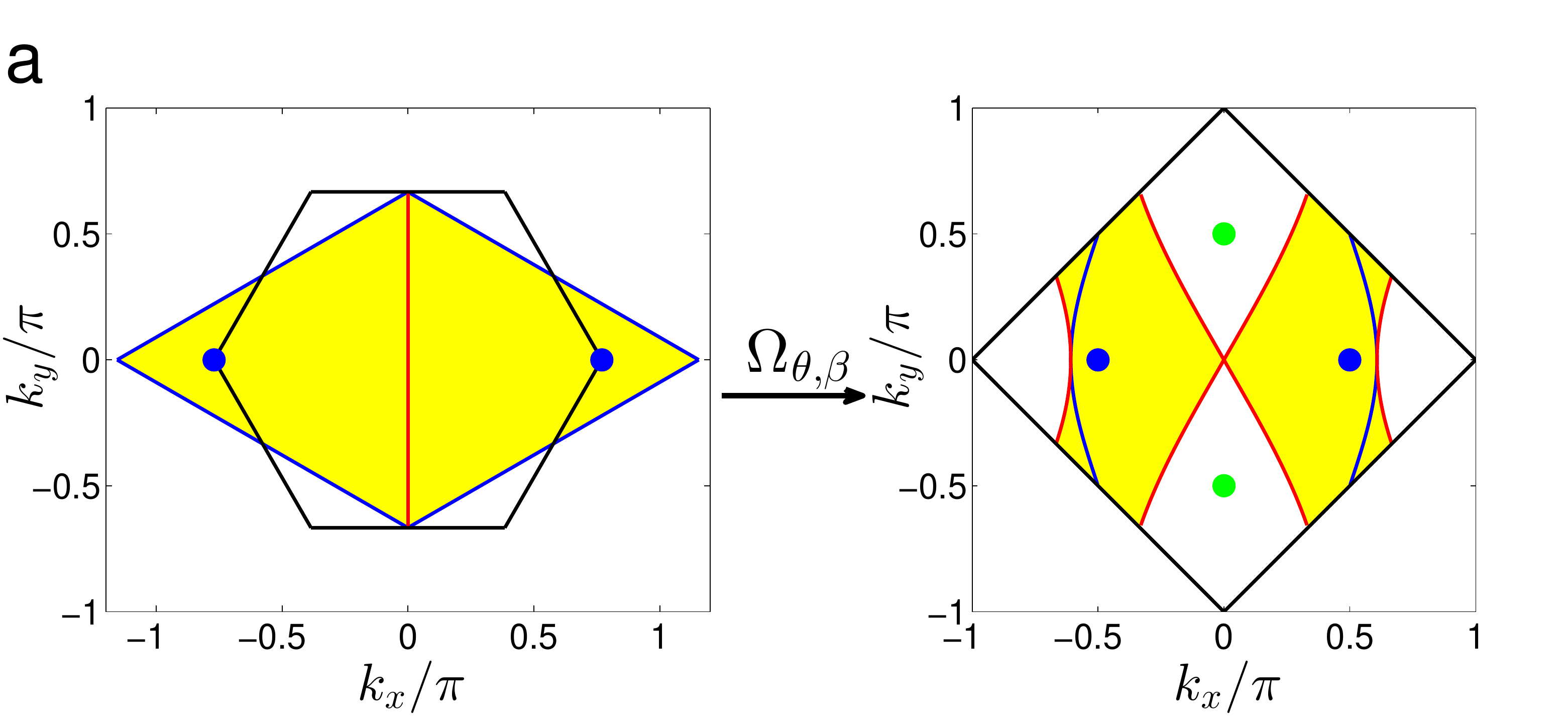}

\includegraphics[width=0.95\columnwidth]{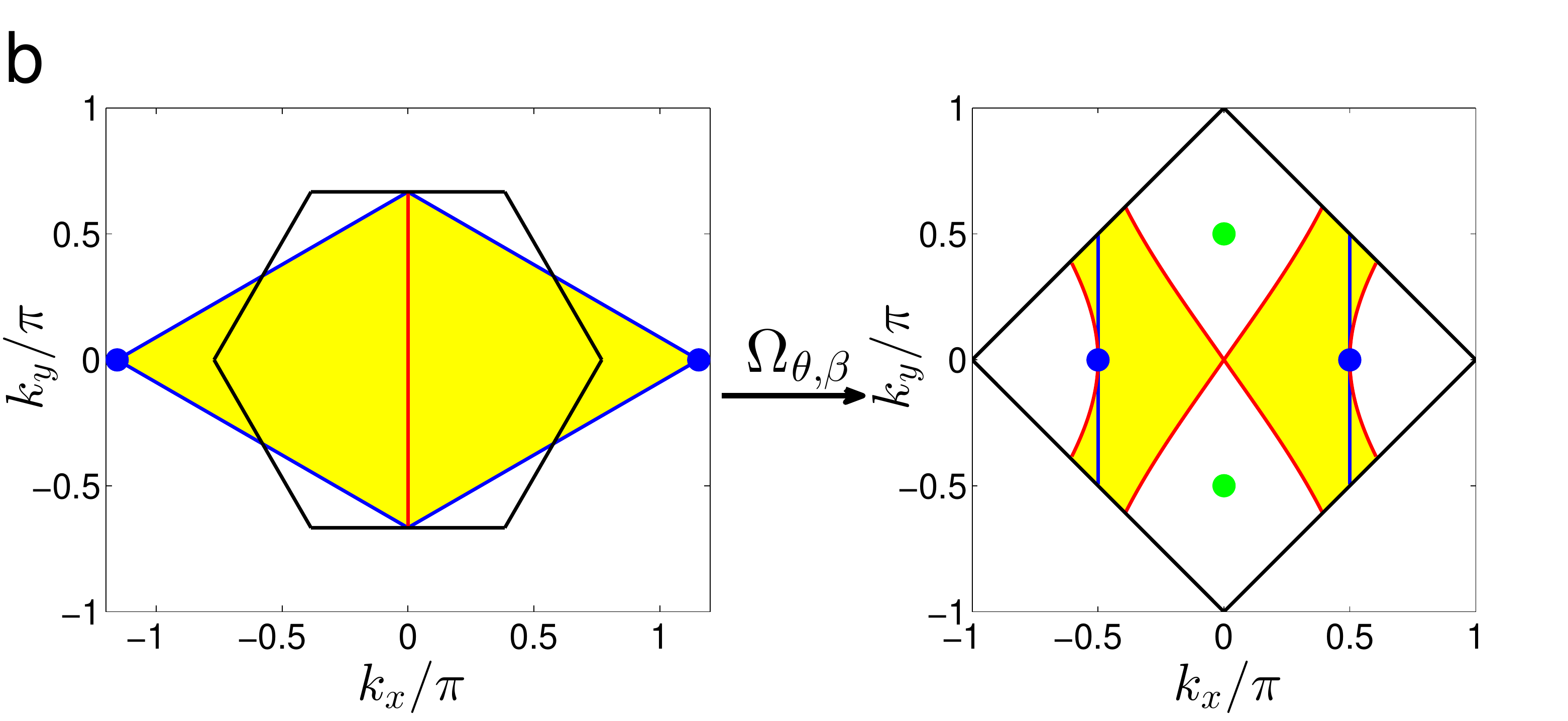}

\includegraphics[width=0.95\columnwidth]{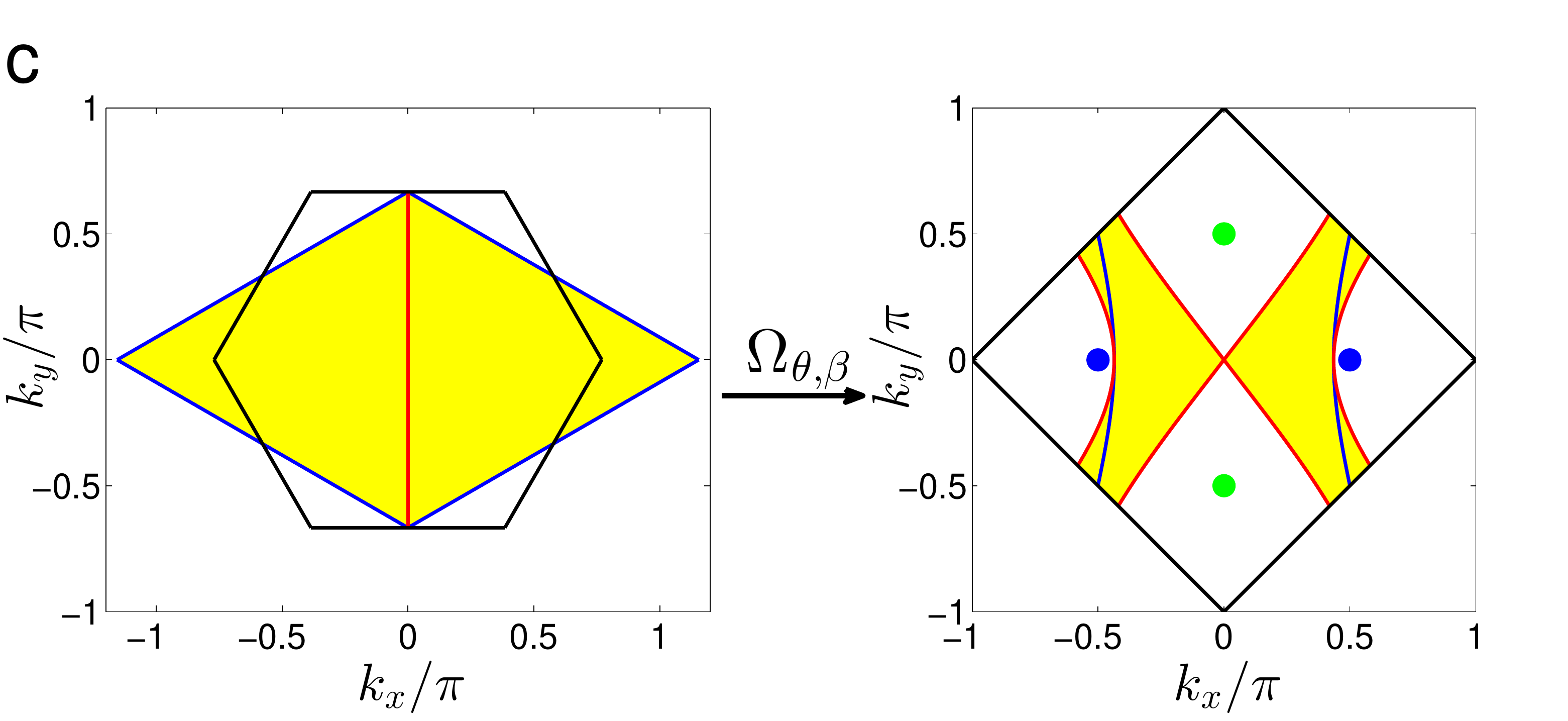}

\caption{ {\bf The mapping from the Brillouin zone of the honeycomb
 lattice into the Brillouin zone of the square lattice.} \textbf{a}, The case with $\beta=1$. \textbf{b}, The case with $\beta=2$. \textbf{c}, The case with $\beta=3$.
 In  the left panels, the  yellow diamond areas represent the Brillouin zone of the honeycomb lattice, which are equivalent to the areas enclosed the black solid
 lines;
 the blue filled circles represent the $\Lambda_{\theta,\beta}$-invariant points $\mathbf{Q}_{1,2}$. In the right panels, the square areas enclosed by the black solid lines represent the
 Brillouin zone of the square lattice; the blue and green filled circles represent the $\Upsilon$-invariant points $\mathbf{M}_{1,2}$ and $\mathbf{M}_{3,4}$, respectively; the yellow areas
 are the image of the mapping $\Omega_{\theta,\beta}$ for the Brillouin zone of the honeycomb lattice.
 The mapping  $\Omega_{\theta,\beta}$ concretely map  the blue filled circles, the blue and red lines   in left panels   into the blue filled circles, the blue
  and red lines in the right panels,
 respectively.    }\label{fig2}
\end{figure}

   With the help of the
mapping $\Omega_{\theta,\beta}$, we define an hidden symmetry
operator of the honeycomb lattice as $\Lambda_{\theta,\beta}=
\Omega_{\theta,\beta}^{-1}\circ\Upsilon\circ\Omega_{\theta,\beta}$,
which means the honeycomb lattice model is invariant under firstly a
mapping into the square lattice, then a $\Upsilon$-transformation,
and finally an inverse mapping back to the honeycomb lattice.
Therefore, the operation $\Lambda_{\theta,\beta}$ can be considered
as a self-mapping of the honeycomb lattice model as
\begin{eqnarray*}
\Lambda_{\theta,\beta}:(\mathbf{k}, {\cal H}_h(\mathbf{k}),
\Psi_{h,\mathbf{k}}(\mathbf{r}))\mapsto (\mathbf{k}', {\cal
H}_h(\mathbf{k}'), {\Psi'_{h,\mathbf{k}'}}(\mathbf{r}))
\end{eqnarray*}
Applying this operator to the wave function, we obtain $
\Lambda_{\theta,\beta}\Psi_{h,\mathbf{k}}(\mathbf{r})={\Psi'_{h,\mathbf{k}'}}(\mathbf{r})
$, where the final momentum is
$\mathbf{k}'=(-k_x-\Delta_x(\mathbf{k})-\Delta_x({\mathbf{k}'}),
-k_y+\pi-\Delta_y(\mathbf{k})-\Delta_y({\mathbf{k}'}))$, with
$\Delta_x(\mathbf{k})=K_x-k_x$ and $\Delta_y(\mathbf{k})=K_y-k_y$
being the shift of the wave vector $\mathbf{k}$ due to the mapping
  $\Omega_{\theta,\beta}$.
If $\mathbf{k}'=\mathbf{k}+\mathbf{K}^h_m$ with $\mathbf{K}_m^h$
being the reciprocal lattice vector of the honeycomb lattice, then
$\mathbf{k}$ is a $\Lambda_{\theta,\beta}$-invariant point. For the
honeycomb lattice model, the $\Lambda_{\theta,\beta}$-invariant
points in the Brillouin zone are $\mathbf{Q}_{1,2}=(\pm
\arccos(-\beta/2)/\cos\theta,0)$.

It is straightforward to verify that
$\Lambda_{\theta,\beta}^2=\Omega_{\theta,\beta}^{-1}\circ\Upsilon^2\circ\Omega_{\theta,\beta}$,
a direct action of which on the Bloch function results in
$\Lambda_{\theta,\beta}^2\Psi_{h,\mathbf{k}}(\mathbf{r})=e^{-2i[k_x+\Delta_x(\mathbf{k})]}\Psi_{h,\mathbf{k}}(\mathbf{r})
$. Substituting  the $\Lambda_{\theta,\beta}$-invariant points
$\mathbf{Q}_{1,2}$ into the above equation, we find
$\Lambda_{\theta,\beta}^2=-1$. Since $\Lambda_{\theta,\beta}$ is an
anti-unitary operator, there must exist band degeneracies  at the
$\Lambda_{\theta,\beta} $-invariant points $\mathbf{Q}_{1,2}$. (For
details see Supplementary Section S-5B). In particular, when
$\theta=\pi/6$ and $\beta=1$, the $\Lambda_{\theta,\beta}
$-protected degenerate points are $\mathbf{Q}_{1,2}=(\pm
4\pi/3\sqrt{3},0)$, just the position of the Dirac points on the
ideal honeycomb lattice, such as graphene. When $\theta=0$ and
$\beta=1$, the $\Lambda_{\theta,\beta} $-protected band degeneracies
occur at $\mathbf{Q}_{1,2}=(\pm 2\pi/3,0)$, which correspond to the
locations of the Dirac points on the brick-wall lattice.

\vspace{0.5cm} \noindent \textbf{Explanation for the moving and
merging of Dirac points and the quantum phase transition}. We have
proved above that the Dirac points on the honeycomb lattice are
protected by the hidden symmetry $\Lambda_{\theta,\beta}$. More
generally, the moving and merging of Dirac points on the  honeycomb
lattice, which has been predicted
theoretically\cite{Hasegawa,Zhu,Wunsch,Pereira,Montambaux2} and
observed experimentally\cite{Tarruell}, can also be explained by the
hidden symmetry. Since the hidden symmetry operator
$\Lambda_{\theta,\beta}$ contains the parameters $\theta$ and
$\beta$, the locations of the $\Lambda_{\theta,\beta}$-protected
Dirac points,
$\mathbf{Q}_{1,2}=(\pm\arccos(-\beta/2)/\cos\theta,0)$,
 are also functions of the two parameters.
As the hopping amplitude ratio $\beta$ starts to increase from 1,
the Dirac points move away from each other. When $\beta$ reaches
$2$, two Dirac points merge into a single one at the corner of the
Brillouin zone. If $\beta$ increases further, there is no solution
to  the $\Lambda_{\theta,\beta} $-invariant points, thereby the
Dirac points vanish, with a gap opening simultaneously. As a result,
$\beta=2$ is the critical point of the quantum phase transition.

 We can interpret the above conclusion
in a more intuitive way by mapping the Brillouin zone of the
honeycomb lattice into that of the square lattice, as shown in
Fig.\ref{fig2}. It turns out that such a mapping is not surjective,
 which means that the image of
the Brillouin zone of the honeycomb lattice  is   part of the
Brillouin zone of the square lattice. In the parameter interval of
$0<\beta<2$, the image covers the $\Upsilon$-protected degenerate
points $\mathbf{M}_{1,2}$ in the Brillouin zone of the square
lattice, as shown in Fig.\ref{fig2}a. Thus, there always exist two
points $\mathbf{Q}_{1,2}$ in the Brillouin zone of the honeycomb
lattice  mapping into the $\Upsilon$-protected degenerate points
$\mathbf{M}_{1,2}$ in the Brillouin zone of the square lattice. When
$\beta=2$, the two equivalent points locating at the corners of the
Brillouin zone of the honeycomb lattice  map into the
$\Upsilon$-protected degenerate points $\mathbf{M}_{1,2}$, as shown
in Fig.\ref{fig2}b. Therefore, the Dirac points on the honeycomb
lattice  merge. When $\beta>2$, the image of the Brillouin zone of
the honeycomb lattice    can not cover the $\Upsilon$-protected
degenerate points $\mathbf{M}_{1,2}$, as shown in Fig.\ref{fig2}c.
Therefore, there is no point in the Brillouin zone of the honeycomb
lattice  mapping into the $\Upsilon$-protected degenerate points
$\mathbf{M}_{1,2}$. Thus, the Dirac points disappear and a gap
opens.

\vspace{0.5cm} \noindent \textbf{Discussion}.

 In summary, we have found a hidden symmetry on
the honeycomb lattice and proved that the hidden symmetry protects
the Dirac points on the honeycomb lattice. The hidden symmetry
evolves along with the parameters, such as the bond angle $\theta$
and the hopping amplitude ratio $\beta$, which provides a perfect
explanation on the moving and merging of the Dirac points and the
quantum phase transition on the honeycomb lattice. Our research
unfolds a new perspective on the symmetry protected band degeneracy,
which is totally different from the conventional ones, such as the
band degeneracy protected by point groups or time reversal symmetry.
such novel hidden symmetry can greatly enrich and deepen our
understanding of the band degeneracy, which will have important
applications in modern condensed matter physics, especially, in the
topics of Dirac (Weyl) semimetals and other topological semimetals.

\vspace{0.5cm} \noindent \textbf{Methods}

\noindent \textbf{The transformation of the wave vectors under the
action of the operator $\Upsilon$.} The square lattice is invariant
under the action of the operator $\Upsilon$, which is anti-unitary.
We suppose that the Bloch functions of the square lattice model have
the form as
\begin{eqnarray}
\Psi_{s,\mathbf{k}}=\left(\begin{matrix}u^{s}_{1,\mathbf{k}}(\mathbf{r})\cr
u^{s}_{2,\mathbf{k}}(\mathbf{r})\end{matrix}\right)e^{i\mathbf{k}\cdot\mathbf{r}}
\label{Bloch}
\end{eqnarray}
where
$u^s_{i,\mathbf{k}}(\mathbf{r})=u^s_{i,\mathbf{k}}(\mathbf{r}+\mathbf{R}_n)$
with $i=1,2$.
 Then, the hidden symmetry operator $\Upsilon$ acts on the
Bloch functions
 as follows
\begin{eqnarray}
{\Psi'_{s,\mathbf{k}'}}(\mathbf{r})
&=&\Upsilon\Psi_{s,\mathbf{k}}(\mathbf{r})\nonumber\\
&=&\left(\begin{matrix}u^{s*}_{2,\mathbf{k}}(\mathbf{r}-\hat{x})e^{ik_x}\cr
u^{s*}_{1,\mathbf{k}}(\mathbf{r}-\hat{x})e^{ik_x}\end{matrix}\right)e^{-i[k_xx+(k_y-\pi)y]}
\label{Bloch2}
\end{eqnarray}
Because $\Upsilon$  is the symmetry operator for the square lattice,
${\Psi'_{s,\mathbf{k}'}}(\mathbf{r})$ must be a Bloch   function of
the square lattice model. Comparing equation (\ref{Bloch2}) with
equation (\ref{Bloch}), we have
\begin{eqnarray}
\Upsilon:(k_x,k_y)\mapsto(k_x',k_y')= (-k_x,-k_y+\pi)
\end{eqnarray}
which can be regarded as the transformation of the wave vector under
the action of the  operator $\Upsilon$.

\section*{Acknowledgement}
\noindent
 We thank S.-L. Zhu  for helpful discussions. This work was
supported by the National Natural Science Foundation of China under
Grants No. 11274061 and No. 11004028.

\section*{Author contributions}
\noindent J.M.H conceived and supervised the project. J.M.H. and
W.C. made the calculations and wrote the paper.

\section*{Additional Information}
\noindent
 {\bf Supplementary
Information} accompanies this paper.
\\
\\
\noindent{\bf Competing financial interests:}  The authors declare
no competing financial interests.


\clearpage

\renewcommand{\thesection}{S-\arabic{section}}
\setcounter{section}{0}  
\renewcommand{\theequation}{S\arabic{equation}}
\setcounter{equation}{0}  
\renewcommand{\thefigure}{S\arabic{figure}}
\setcounter{figure}{0}  

\onecolumngrid
\
\\
\begin{center}
 {\textbf{\Large    Hidden symmetry and protection of Dirac points  on the
honeycomb lattice---Supplementary  information }}
 \end{center}
 \centerline{Jing-Min Hou$^1$  and Wei Chen$^2$}
 \centerline{\it $^1$Department of Physics,
Southeast University, Nanjing  211189, China} \centerline{\it
$^2$College of Science, Nanjing University of Aeronautics and
Astronautics, Nanjing 210016, China}

\section{The derivation of the Bloch Hamiltonian and the dispersion relation }

\subsection{The honeycomb lattice}

For the general honeycomb lattice with the bond angle $\theta$, the
tight-binding Hamiltonian can be written as,
\begin{eqnarray}
  H_h&=&-\sum_{i\in A} [ t_1 a^\dag_{i}
b_{i+\hat{d}_1}  +  t_1 a^\dag_{i} b_{i+\hat{d}_2}
    + t_2a^\dag_{i}b_{i+\hat{d}_3}] + H.c.
    \label{THh}
\end{eqnarray}
 where $a_i  $ is the
annihilation operator that destructs  a particle in the Wannier
state located at the site $i$ in sublattice $A$ and  $b_j  $ is the
annihilation operator that destructs  a particle in the Wannier
state located at the site $j$ in sublattice $B$;
 the subscript $i\equiv (i_x,i_y)$ is the coordinate for the
lattice sites; $\hat{d}_1=(\cos\theta, \sin\theta),
\hat{d}_2=(-\cos\theta, \sin\theta), \hat{d}_3=(0,-1)$
 represent the
unit vectors between the two nearest lattice sites;  $t_1$ and $t_2$
are the amplitudes of hopping as shown  in Fig.1 a in the main text.
 We take the Fourier's transformation to the annihilation operators as
\begin{eqnarray}
{a}_{\mathbf{k}}&=&\frac{1}{\sqrt{N}}\sum_{i}
  {a}_{i}e^{-i\mathbf{k}\cdot\mathbf{R}^A_i},\label{fta}\\
  {b}_{\mathbf{k}}&=&\frac{1}{\sqrt{N}}\sum_{j }
  {b}_{j}e^{-i\mathbf{k}\cdot\mathbf{R}^B_j},\label{ftb}
\end{eqnarray}
where $\mathbf{R}_i^A$ and $\mathbf{R}^B_j$ represent the positions
of lattice sites in sublattice lattice $A$ and $B$, respectively.
Substituting Eqs.(\ref{fta}) and (\ref{ftb}) into equation
(\ref{THh}), we obtain
\begin{eqnarray}
  H_h&=&- \sum_{\mathbf{k}} [ 2t_1\cos(\cos\theta k_x) e^{i\sin\theta k_y} a^\dag_{\mathbf{k}}
b_{\mathbf{k}}
   +t_2e^{-ik_y}a^\dag_{\mathbf{k}}b_{\mathbf{k}} ]+H.c.
\end{eqnarray}
We define the two-component annihilation operator as $
 {\eta}_{\mathbf{k}}\equiv  [
 {a}_{\mathbf{k}},
 {b}_{\mathbf{k}} ]^T$ and the Hamiltonian can be written as $H_h=\sum_\mathbf{k}\eta_\mathbf{k}^\dag
 \mathcal{H}_h(\mathbf{k})\eta_\mathbf{k}$.
Here, $\mathcal{H}_h(\mathbf{k})$ is the Bloch Hamiltonian of the
honeycomb lattice model   for the wave vector $\mathbf{k}$ as
\begin{eqnarray}
{\cal H}_h(\mathbf{k})&=&-[ 2t_1\cos(\cos\theta k_x)\cos(\sin\theta
k_y) +t_2\cos k_y]\sigma_x+[2t_1\cos(\cos\theta k_x)\sin(\sin\theta
k_y)-t_2\sin k_y]\sigma_y\label{BHh2}
\end{eqnarray}
where $\sigma_x$ and $\sigma_y$ are the Pauli matrices.  This Bloch
Hamiltonian is equation   (1) in the main text.  The corresponding
dispersion relation is
\begin{eqnarray}
E_h(\mathbf{k})=\pm\sqrt{4t_1^2\cos^2(\cos\theta
k_x)+4t_1t_2\cos(\cos\theta k_x)\cos[(1+\sin\theta)k_y]+t_2^2}
\end{eqnarray}

For the ideal honeycomb lattice, $\theta=\pi/6$, such as graphene,
the Bloch Hamiltonian is
\begin{eqnarray}
{\cal H}_h(\mathbf{k})&=&-[ 2t_1\cos( {\sqrt{3}} k_x/2)\cos( k_y/2)
+t_2\cos k_y]\sigma_x+[2t_1\cos( {\sqrt{3}} k_x/2)\sin(
k_y/2)-t_2\sin k_y]\sigma_y
\end{eqnarray} and
the dispersion relation is
\begin{eqnarray}
E_h(\mathbf{k})=\pm\sqrt{4t_1^2\cos^2(\sqrt{3}
k_x/2)+4t_1t_2\cos(\sqrt{3}k_x/2)\cos(3k_y/2)+t_2^2}
\end{eqnarray}

The honeycomb lattice with $\theta=0$ is the brick-wall lattice.
Substituting $\theta=0$ into equation (\ref{BHh2}), we obtain the
Bloch Hamiltonian for the brick-wall lattice as
\begin{eqnarray}
{\cal H}_b(\mathbf{k})=-[ 2t_1\cos k_x +t_2\cos k_y]\sigma_x
-t_2\sin k_y\sigma_y\label{SHB}
\end{eqnarray}
 The corresponding
dispersion relations is
\begin{eqnarray}
E_b(\mathbf{k})=\pm\sqrt{4t_1^2\cos^2k_x+4t_1t_2\cos k_x\cos
k_y+t_2^2}
\end{eqnarray}

\subsection{The square lattice}
For the square lattice with a hopping-accompanying $\pi$ phase, the
tight-binding Hamiltonian can be written as
\begin{eqnarray}
  H_s&=&-\sum_{i\in A} [ t_x  a^\dag_{i}
b_{i+\hat{x}}  +  t_x a^\dag_{i} b_{i-\hat{x}}
 +t_ye^{-i\pi} a^\dag_{i} b_{i+\hat{y}} +t_ya^\dag_{i}b_{i-\hat{y}}] +H.c.
\end{eqnarray}
where $\hat{x}$ and $\hat{y}$ represent the unit vectors in the $x$
and $y$ directions, respectively; $t_x$ and $t_y$ are the amplitudes
of hopping along the $x$ and $y$ directions, respectively. Taking
the Fourier's transformation, we obtain the Hamiltonian as
\begin{eqnarray}
  H_s&=&- \sum_{\mathbf{k}} [ 2t_x\cos k_x a^\dag_{\mathbf{k}}
b_{\mathbf{k}}
  -2t_yi\sin k_y a^\dag_{\mathbf{k}} b_{\mathbf{k}}]+H.c.
\end{eqnarray}
We rewritten the Hamiltonian as
$H_s=\sum_\mathbf{k}\eta_\mathbf{k}^\dag
 \mathcal{H}_s(\mathbf{k})\eta_\mathbf{k}$ with  $
 {\eta}_{\mathbf{k}}\equiv  [
 {a}_{\mathbf{k}},
 {b}_{\mathbf{k}} ]^T$.
 Then, the Bloch  Hamiltonian for the square lattice model is
 \begin{eqnarray}
{\cal H}_s(\mathbf{k})=- 2t_x\cos k_x \sigma_x -2t_y\sin k_y\sigma_y
\end{eqnarray}
The corresponding dispersion relation is
\begin{eqnarray}
E_s(\mathbf{k})=\pm\sqrt{4t_x^2\cos^2 k_x+ 4t_y^2 \sin^2 k_y}
\end{eqnarray}

\begin{figure*}[ht]
\includegraphics[width=0.3840\columnwidth]{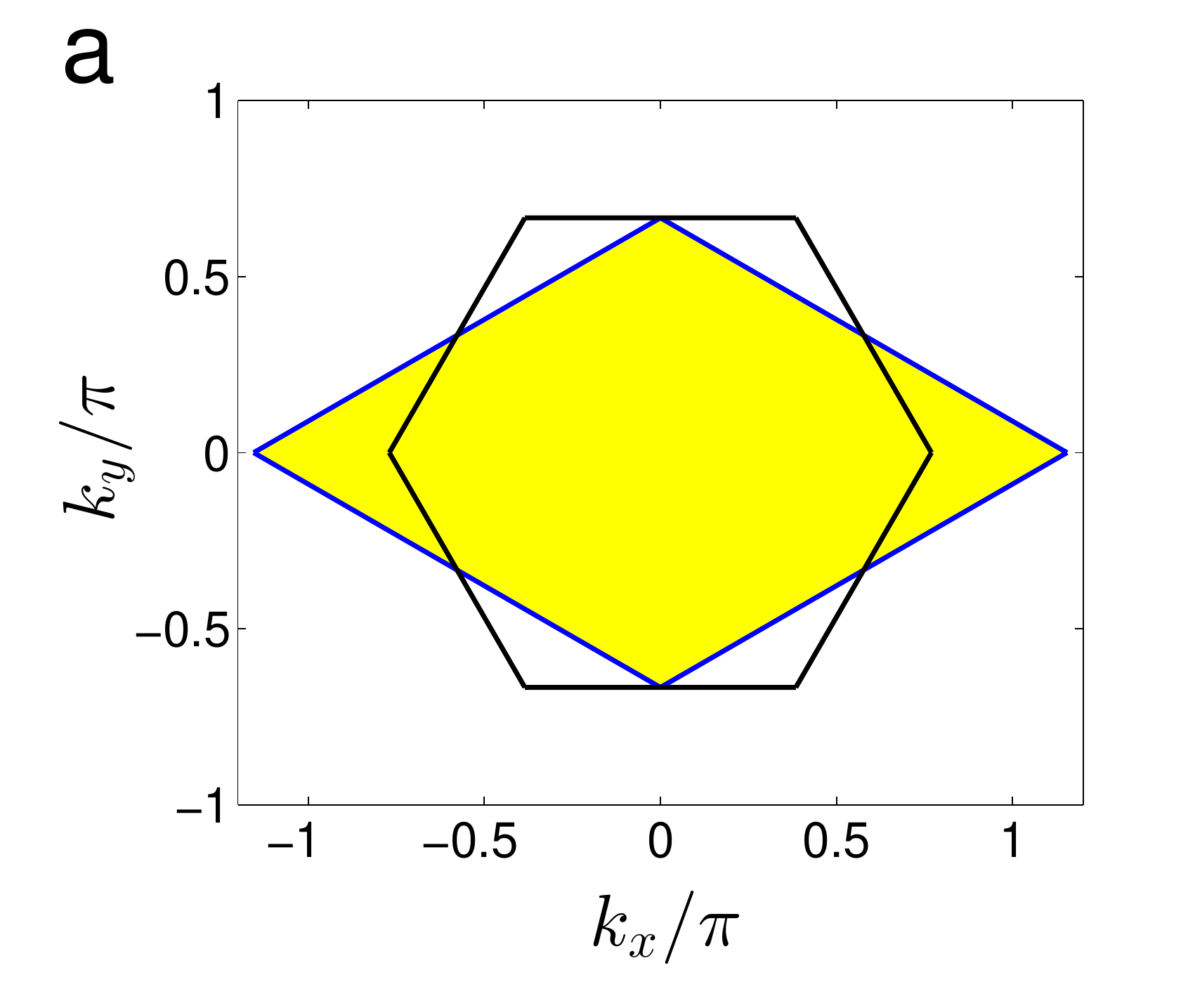}
\includegraphics[width=0.32\columnwidth]{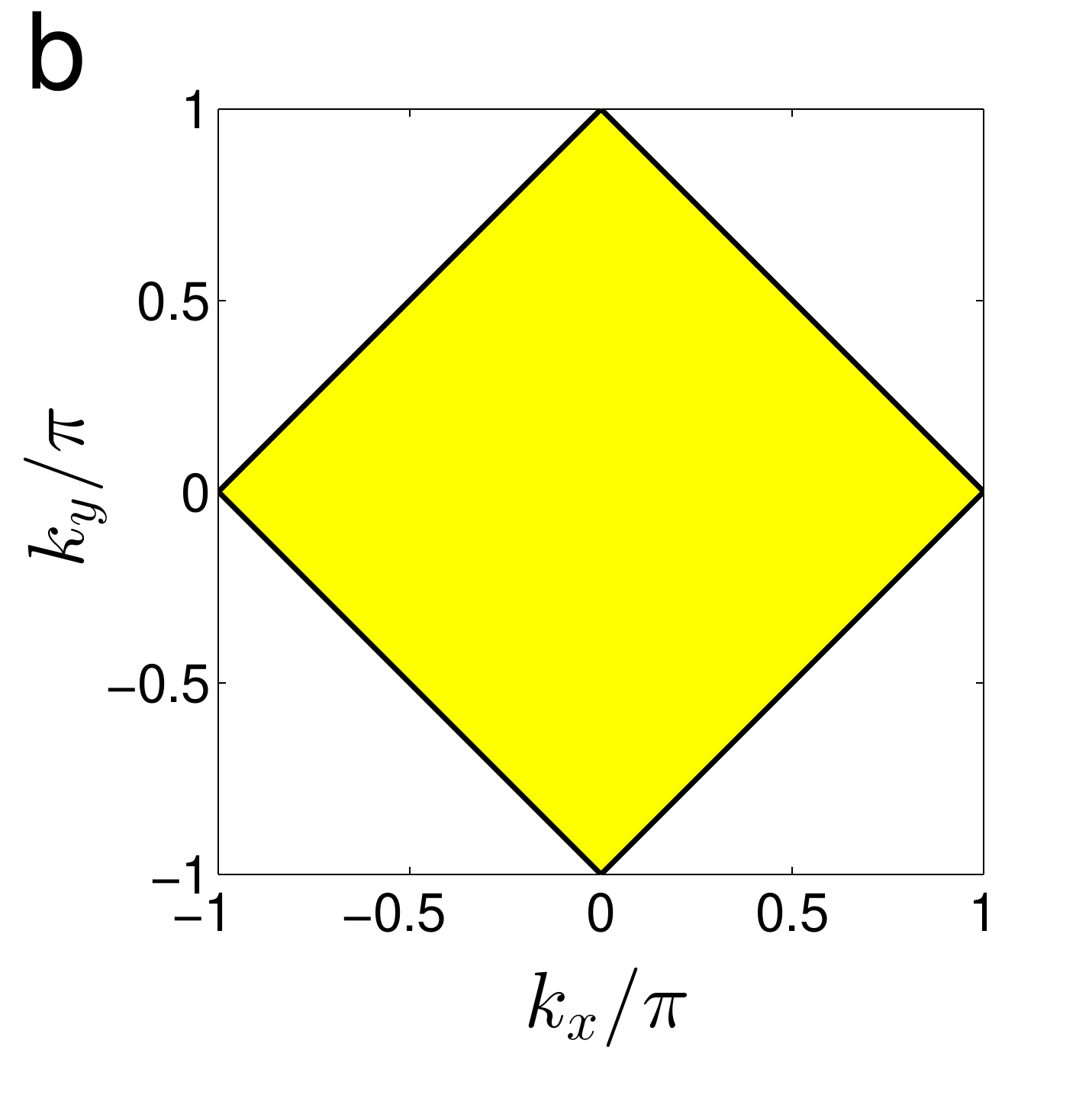}
\caption{{\bf The Brillouin zones.} {\bf  a}, The Brillouin zone of
the honeycomb lattice. The Brillouin zone of the honeycomb lattice
can be represented by the hexagon area enclosed by the black solid
lines, equivalently, it can also be represented by the yellow
diamond.  {\bf b}, The Brillouin zone of the brick-wall lattice and
the square lattice.} \label{figS1}
\end{figure*}
\section{The Brillouin zone}

For all the   lattices, we assume that the bond length $a=1$.

For the general honeycomb lattice with the bond angle $\theta$, the
primitive lattice vectors are
$\mathbf{a}_1=(\cos\theta,1+\sin\theta)$ and
$\mathbf{a}_2=(\cos\theta,-1-\sin\theta)$.
 The primitive reciprocal lattice vectors are
 $\mathbf{b}_1=({\pi}/{\cos\theta},{\pi}/({1+\sin\theta}))$ and
 $\mathbf{b}_2=({\pi}/{\cos\theta},-{\pi}/({1+\sin\theta}))$.
For $\theta=\pi/6$ case, such as graphene, the symmetric Brillouin
zone is hexagon, i.e., the area enclosed by the black lines  in
Fig.\ref{figS1} a. An alternative Brillouin zone equivalent to the
symmetric Brillouin zone is a diamond, i.e., the yellow shaded area
in Fig.\ref{figS1} a.  In our work, for convenience, we always use
the diamond Brillouin zone for the honeycomb lattice.

 For  and the square lattice,
 the primitive
lattice vectors are $\mathbf{a}_1=(1,1)$ and $\mathbf{a}_2=(1,-1)$.
 The primitive reciprocal lattice vectors are
 $\mathbf{b}_1=(\pi,\pi)$ and
 $\mathbf{b}_2=(\pi,-\pi)$. The   square
 lattice has a square Brillouin zone as shown  in
 Fig.\ref{figS1} b.

The brick-wall lattice  can be considered a special honeycomb
lattice with the bond angle $\theta=0$. The primitive lattice
vectors become $\mathbf{a}_1=(1,1)$ and $\mathbf{a}_2=(1,-1)$.
 The primitive reciprocal lattice vectors are
 $\mathbf{b}_1=(\pi,\pi)$ and
 $\mathbf{b}_2=(\pi,-\pi)$.  The corresponding
Brillouin zone turns into a square, which is the same with that of
the square lattice as shown in Fig.\ref{figS1} b.

\section{The hidden symmetry of the square lattice}
\begin{figure*}[h]
\includegraphics[width=0.8\columnwidth]{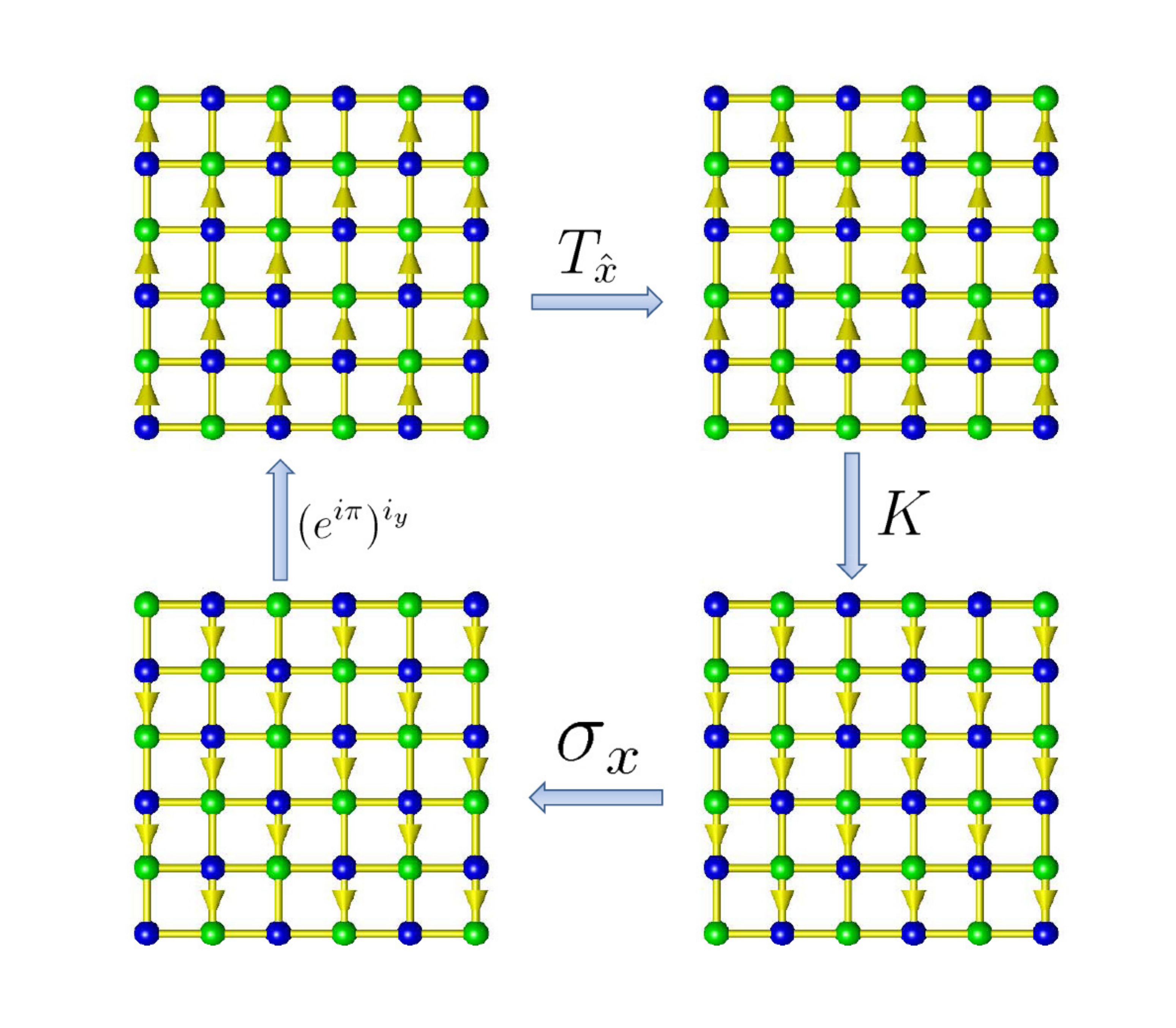}
\caption{ {\bf Schematic of the invariance of the square lattice
under the action of the hidden symmetry $\Upsilon$}. The hidden
symmetry consists of the translation transformation $T_{\hat{x}}$,
the complex conjugation $K$, the sublattice exchange $\sigma_x$ and
the local $U(1)$ gauge transformation $(e^{i\pi})^{i_y}$ in order.
Here, the arrows represent a hopping-accompanying $\pi$ phase. }
\label{symmetry}
\end{figure*}

The square lattice with a hopping-accompanying $\pi$ phase
  respects a hidden symmetry, which is defined
as
\begin{eqnarray*}
\Upsilon=(e^{i\pi})^{i_y}\sigma_x K T_{\hat{x}}
\end{eqnarray*}
where $T_{\hat{x}}$, $K$, $\sigma_x$, and $(e^{i\pi})^{i_y}$
represent a translation along the $x$ direction by a unit vector,
the complex conjugation, the sublattice exchange, and a local $U(1)$
gauge transformation, respectively. From Fig.\ref{symmetry}, we can
find the square lattice with a hopping-accompanying $\pi$ phase is
invariant under the actions of $T_x$, $K$, $\sigma_x$, and
$(e^{i\pi})^{i_y}$ in order. Thus, we conclude that the square
lattice with a hopping-accompanying $\pi$ phase respects the hidden
symmetry $\Upsilon$.

\section{Explanation for the mapping $\Omega_{\theta,\beta}$}

We can interpret the mapping $\Omega_{\theta,\beta}$ in an intuitive
way. To this end, we divide it into two mappings in order as
$\Omega_{\theta,\beta}=\omega_{2,\beta}\circ\omega_{1,\theta}$.
Here, $\omega_{1,\theta}$ is the mapping from the general honeycomb
lattice model with the bond angle $\theta$ into the brick-wall
lattice model and $\omega_{2,\beta}$ is the mapping from the
brick-wall lattice model with the hopping amplitude ratio $\beta$
into the square lattice model. In the following, we explain  the two
mappings in details.

\subsection{The mapping $\omega_{1,\theta}$}

\begin{figure*}[ht]
\includegraphics[width=0.67\columnwidth]{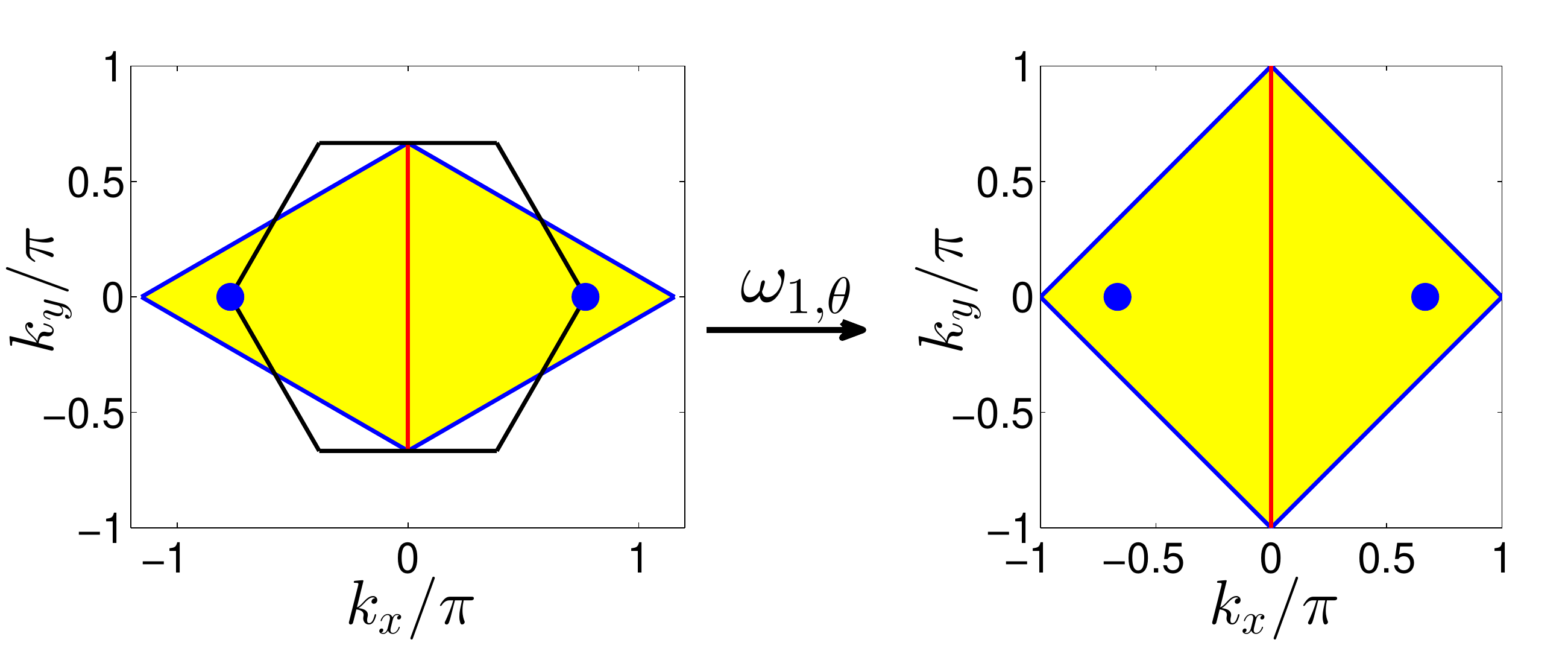}
\caption{{\bf The mapping from the Brillouin zone of the honeycomb
lattice   into the Brillouin zone of the brick-wall lattice.} The
left and right panels show the Brillouin zones for the honeycomb
lattice  with $\theta=\pi/6$ and the brick-wall lattices,
respectively.  Concretely, the blue lines, the red line, and the
blue filled circles (the $\Lambda_{\theta,\beta}$-invariant points)
in the left panel map into the blue lines, the red line, and the
blue filled circles in the right panel, respectively.} \label{map1}
\end{figure*}

The general honeycomb lattice model with the bond angle $\theta$ is
equivalent with the brick-wall lattice model in some sense. To
manifest this equivalence, we define a mapping $\omega_{1,\theta}$
from the general honeycomb lattice model to the brick-wall lattice
model as
\begin{eqnarray*}
\omega_{1,\theta}: (\mathbf{k}, {\cal H}_h(\mathbf{k}),
\Psi_{h,\mathbf{k}}(\mathbf{r}))\mapsto (\mathbf{p}, {\cal
H}_b(\mathbf{p}), \Psi_{b,\mathbf{p}}(\mathbf{r}))
\end{eqnarray*}
where  $\Psi_{h,\mathbf{k}}(\mathbf{r})$ and
$\Psi_{b,\mathbf{p}}(\mathbf{r})$ are the Bloch functions of the
honeycomb lattice model and the brick-wall lattice model,
respectively.  To find the explicit form of the mapping, we take a
transformation to the Bloch Hamiltonian (equation 1 in the main
text) as ${\cal H}_h'(\mathbf{k})=S_{\mathbf{k}}{\cal
H}_h(\mathbf{k})S_{\mathbf{k}}^{-1} $, where $S_\mathbf{k}$  is the
transformation matrix defined as
\begin{eqnarray}
S_\mathbf{k}=\frac{1}{2}\left(\begin{matrix}1+\frac{(A_\mathbf{k}+iB_\mathbf{k})(C_\mathbf{k}-iD_\mathbf{k})}{\sqrt{A_\mathbf{k}^2+B_\mathbf{k}^2}}&
\frac{-(A_\mathbf{k}-iB_\mathbf{k})+(C_\mathbf{k}-iD_\mathbf{k})}{\sqrt{A_\mathbf{k}^2+B_\mathbf{k}^2}}\cr
\frac{(A_\mathbf{k}+iB_\mathbf{k})-(C_\mathbf{k}+iD_\mathbf{k})}{\sqrt{A_\mathbf{k}^2+B_\mathbf{k}^2}}&
1+\frac{(A_\mathbf{k}-iB_\mathbf{k})(C_\mathbf{k}+iD_\mathbf{k})}{\sqrt{A_\mathbf{k}^2+B_\mathbf{k}^2}}\end{matrix}\right)
\end{eqnarray}
for $A_\mathbf{k}^2+B_\mathbf{k}^2\neq 0$. Here
$A_\mathbf{k}=2t_1\cos(\cos\theta k_x)\cos(\sin\theta k_y) +t_2\cos
k_y$, $B_\mathbf{k}=t_2\sin k_y-2t_1\cos(\cos\theta
k_x)\sin(\sin\theta k_y)$, $C_\mathbf{k}=2t_1\cos (\cos\theta k_x)
+t_2\cos [(1+\sin\theta)k_y]$ and $D_\mathbf{k}=t_2\sin
[(1+\sin\theta)k_y]$, which satisfy the identity
$A_\mathbf{k}^2+B_\mathbf{k}^2=C_\mathbf{k}^2+D_\mathbf{k}^2$.
 When $A^2+B^2=0$, the transformation matrix $S_\mathbf{k}$ is
 ill-defined. For the continuity of the mapping, when $A^2+B^2=0$, we take the limit
 as the definition of $S_\mathbf{k}$. We  then obtain
\begin{eqnarray}
{\cal H}'_h(\mathbf{k})=-\{[ 2t_1\cos (\cos\theta k_x) +t_2\cos
[(1+\sin\theta)k_y]\}\sigma_x -t_2\sin [(1+\sin\theta)k_y]\sigma_y
\end{eqnarray}
Substituting $ k_x=p_x/\cos\theta$ and $k_y=p_y/(1+\sin\theta)$ into
$\mathcal{H}'_h(\mathbf{k})$, we obtain
\begin{eqnarray}
{\cal H}_b(\mathbf{p})=-[ 2t_1\cos p_x +t_2\cos p_y]\sigma_x
-t_2\sin p_y\sigma_y \label{BHb2}
\end{eqnarray}
which is just  the Bloch Hamiltonian (equation \ref{SHB}) of the
brick-wall lattice model. The mapping $\Omega_{1,\theta}$ has the
effect on the Bloch functions and the wave vectors as $
\omega_{1,\theta}\Psi_{h,\mathbf{k}}(\mathbf{r})=\Psi_{b,\mathbf{p}}(\mathbf{r})
$ and $\omega_{1,\theta}:(k_x,k_y)\mapsto (p_x,p_y)=(\cos\theta k_x,
(1+\sin\theta)k_y)$. This mapping   is one-to-one and surjective.
Thus, we can regard this mapping as a kind of equivalence. The
explicit form of the mapping
 depends on the bond angle $\theta$. When
$\theta=0$, this mapping is an identity mapping.

The mapping $\omega_{1,\theta}$ gives a one-to-one correspondence
between the Brillouin zones of the honeycomb lattice   and the
brick-wall lattice.    That is to say, for some wave vector
$\mathbf{k}$ in the Brillouin zone of the honeycomb lattice, the
Bloch Hamiltonian $\mathcal{H}_h(\mathbf{k})$ and its Bloch
functions $\Psi_{h,\mathbf{k}}(\mathbf{r})$, there correspondingly
exist  a wave vector $\mathbf{p}$ in the Brillouin zone of the
brick-wall lattice, the Bloch Hamiltonian
$\mathcal{H}_b(\mathbf{p})$, and Block functions
$\Psi_{b,\mathbf{p}}(\mathbf{r})$. The mapping from the Brillouin
zone of the honeycomb lattice  into that of the brick-wall lattice
is schematically shown in Fig.\ref{map1}.

\subsection{The mapping $\omega_{2,\beta}$}

\begin{figure*}[ht]
\includegraphics[width=1\columnwidth]{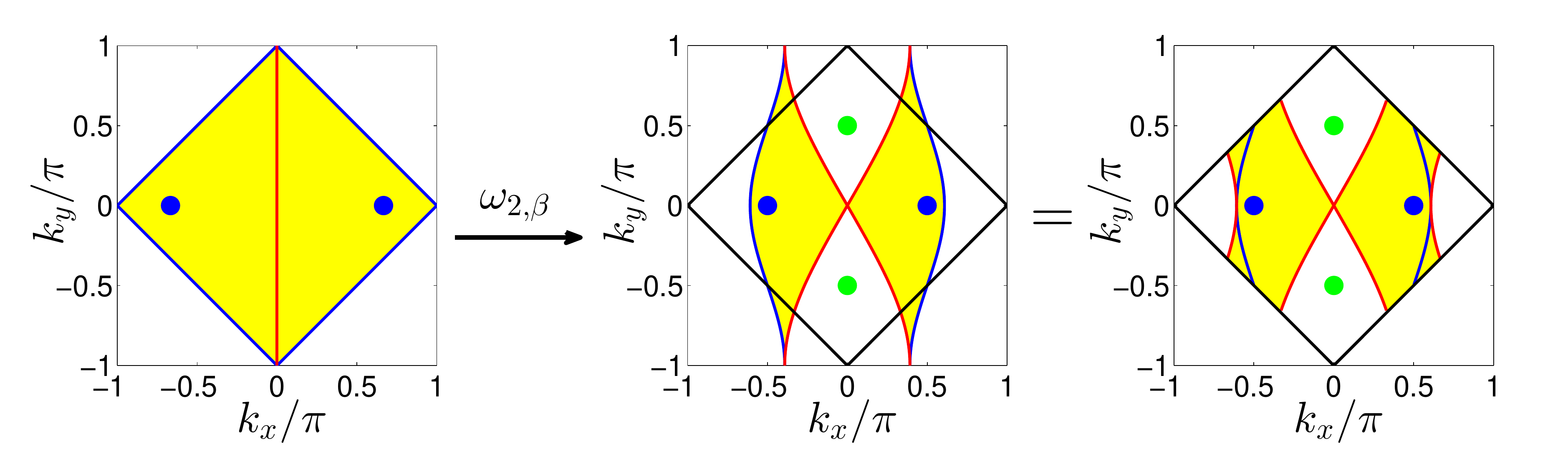}
\caption{{\bf The mapping from the Brillouin zone of the brick-wall
lattice  into the Brillouin zone of the square lattice.} The left
panel shows the Brillouin zone of the  brick-wall lattices; the
middle panel shows the image of the mapping $\omega_{2,\beta}$ for
the Brillouin zone of the brick-wall lattice in the momentum space
of the square lattice; the right panel shows the image of the
mapping $\omega_{2,\beta}$ for the Brillouin zone of the brick-wall
lattice restricted in  the Brillouin zone of the square lattice.
Here, the left and right half of the Brillouin zone of the
brick-wall lattice  in the left panel map into the the left and
right yellow shadow areas in the middle panel, which is equivalent
with that in the right panel. The blue and red lines in the left
panel for the brick-wall lattice  map into the blue and red lines in
the middle  panel for  the square lattice, respectively. The blue
filled circles in the left panel map into the degenerate
$\Upsilon$-invariant points (blue filled circles in the middle and
right panels)  in the Brillouin zone of the square lattice. }
\label{map2}

\end{figure*}

Next, we can define a mapping from the brick-wall lattice model to
the square lattice model as
\begin{eqnarray*}
\omega_{2,\beta}: (\mathbf{p}, {\cal H}_b(\mathbf{p}),
\Psi_{b,\mathbf{p}}(\mathbf{r}))\mapsto (\mathbf{K}, {\cal
H}_s(\mathbf{K}), \Psi_{s,\mathbf{K}}(\mathbf{r}))
\end{eqnarray*}
where   $\beta=t_2/t_1$ is the hopping amplitude ratio of the
brick-wall lattice.
  To find the explicit form of this mapping, we
suppose $
\omega_{2,\beta}\Psi_{b,\mathbf{p}}=\Psi_{s,\mathbf{K}}(\mathbf{r})
$ and
\begin{eqnarray*}
K_x&=&
\begin{cases}
-\arccos[\frac{2}{2+\beta}\cos
p_x+\frac{ \beta }{2+\beta}\cos p_y],&\mbox{for}\  k_x\leq0\\
\arccos[\frac{2}{2+\beta}\cos p_x+\frac{ \beta }{2+\beta}\cos p_y],&
\mbox{for} \ k_x\geq0
\end{cases}\\
K_y&=&p_y
\end{eqnarray*}
where $\mathbf{p}$ is a wave vector in the Brillouin zone of the
brick-wall lattice. Through the mapping $\omega_{2,\beta}$, the
Bloch Hamiltonian of the brick-wall lattice (\ref{BHb2}) becomes the
form as follows,
\begin{eqnarray}
{\cal H}_s(\mathbf{K})=-  2t_x\cos K_x \sigma_x -2t_y\sin
K_y\sigma_y
\end{eqnarray}
with $t_y=t_2/2$ and $t_x=t_1+t_2/2$, which is just the Bloch
Hamiltonian of the square  lattice model. The explicit form of this
mapping depends on the hopping amplitude ratio $\beta$.

 This mapping is not
surjective. That is to say, the image of the mapping for the
Brillouin zone of the Brick-wall lattice  just covers   part of the
Brillouin zone of the   square lattice.  The mapping for the wave
vectors is schematically shown in Fig.\ref{map2}. In Fig.\ref{map2},
the left panel shows the Brillouin zone of the brick-wall lattice.
In order to clearly manifest the mapping from the brick-wall lattice
model to the square lattice model, we first map the Brillouin zone
of the brick-wall lattice into the momentum space of the square
lattice, not restricted in the Brillouin zone, as shown in the
middle panel of Fig.\ref{map2}. The image  of the Brillouin zone of
the brick-wall lattice   in the momentum space of the square lattice
  looks like a butterfly. The left and right halves of the
Brillouin zone of the brick-wall lattice map into the left and right
wings of the butterfly, respectively. If we restrict the image of
the mapping in the Brillouin zone of the square lattice, then the
butterfly-like image is equivalent to that as shown in the right
panel of Fig.\ref{map2}.

\section{The proof of band degeneracy at the Dirac points}
\subsection{The square lattice}
$\Upsilon$ is the operator of the hidden symmetry that is respected
by the square lattice considered in the main  text. The
$\Upsilon$-invariant points in the Brillouin zone are
$\mathbf{M}_{1,2}=( \pm\pi/2, 0)$  and
$\mathbf{M}_{3,4}=(0,\pm\pi/2)$.
 Since the square of the hidden symmetry operator $\Upsilon$ is a  translation operator as $\Upsilon^2=T_{2\hat{x}}$, we have
$
\Upsilon^2\Psi_{s,\mathbf{k}}(\mathbf{r})=T_{2\hat{x}}\Psi_{s,\mathbf{k}}(\mathbf{r})=e^{-2ik_x}\Psi_{s,\mathbf{k}}(\mathbf{r})
$.
 We define $(\psi,
\varphi)$ as the inner product of the two wave functions $\psi$ and
$\varphi$. The anti-unitary operator $\Upsilon$ has the property
that $(\Upsilon\psi, \Upsilon\varphi)=(\psi, \varphi)^*=(\varphi,
\psi)$.
 Therefore, at the $\Upsilon$-invariant point $\mathbf{M}_i$ in the Brillouin zone,  we have
\begin{eqnarray}
({\Psi'_{s,\mathbf{M}_i}},\Psi_{s,\mathbf{M}_i})&=&(\Upsilon\Psi_{s,\mathbf{M}_i},\Upsilon{\Psi'_{s,\mathbf{M}_i}})
=(\Upsilon\Psi_{s,\mathbf{M}_i},\Upsilon^2\Psi_{s,\mathbf{M}_i})\nonumber\\
&=&e^{-2iM_{ix}}( {\Psi'_{s,\mathbf{M}_i}},
\Psi_{{s,\mathbf{M}_i}}).\label{prot}
\end{eqnarray}
where $M_{ix}$ is the $x$ component of the $\Upsilon$-invariant
point $\mathbf{M}_i$ and the input of the Bloch functions is omitted
for convenience. For the $\Upsilon$-invariant points
$\mathbf{M}_{1,2}$, where  $M_{ix}=\pm\pi/2$,
  we have $\Upsilon^2=-1$, then we obtain the solution
$({\Psi'_{s,\mathbf{M}_i}},\Psi_{s,\mathbf{M}_i})=0$, that is to
say, ${\Psi'_{s,\mathbf{M}_i}}$ and $\Psi_{s,\mathbf{M}_i }$ are
orthogonal to each other. While, for the $\Upsilon$-invariant points
$\mathbf{M}_{3,4}$, where $M_{ix}= 0$, we have $\Upsilon^2=1$, so
$({\Psi'_{s,\mathbf{M}_i}},\Psi_{s,\mathbf{M}_i})$ is unconstrained
for equation   (\ref{prot}). Therefore, we arrive at the conclusion
that the system must be degenerate at points $\mathbf{M}_{1,2}$
 in the Brillouin zone for the square lattice, which is consistent with the dispersion relation calculated.

\subsection{The honeycomb lattice}

$\Lambda_{\theta,\beta}$ is a self-mapping of the honeycomb lattice
model with the bond angle $\theta$ and  the hopping amplitude ratio
$\beta$. We assume that $\Psi_{h,\mathbf{k}}(\mathbf{r})$ is the
Bloch function of the honeycomb lattice model. Since
$\Lambda_{\theta,\beta}$ is a self-mapping of the honeycomb lattice
model, we have
\begin{eqnarray}
\Psi'_{h,\mathbf{k}'}(\mathbf{r})=\Lambda_{\theta,\beta}\Psi_{h,\mathbf{k}}(\mathbf{r})
\end{eqnarray}
which is also the Bloch function of the honeycomb lattice model.
After the action of the operator $\Lambda_{\theta,\beta}$, the wave
vector $\mathbf{k}$ becomes
\begin{eqnarray}
\mathbf{k}'=(-k_x-\Delta_x(\mathbf{k})-\Delta_x({\mathbf{k}'}),
-k_y+\pi-\Delta_y(\mathbf{k})-\Delta_y({\mathbf{k}'})) \label{trans}
\end{eqnarray}
where
\begin{eqnarray}
\Delta_x(\mathbf{k})&=&
\begin{cases}
-k_x-\mathcal{K}_{\theta,\beta}(\mathbf{k}),&\mbox{for}\  k_x\leq0\\
-k_x+\mathcal{K}_{\theta,\beta}(\mathbf{k}),& \mbox{for} \ k_x\geq0
\end{cases}\\
\Delta_y(\mathbf{k}))&=&\sin\theta k_y
\end{eqnarray}
with
$\mathcal{K}_{\theta,\beta}(\mathbf{k})\equiv\arccos\{\frac{2}{2+\beta}\cos
(\cos\theta k_x)+\frac{ \beta }{2+\beta}\cos [(1+\sin\theta)k_y]\}$.
 If
\begin{eqnarray}
 \mathbf{k}'=\mathbf{k}+\mathbf{K}_m^h \label{invariant}
\end{eqnarray}
  is satisfied, where $\mathbf{K}_m^h=(
{K}^h_{m,x}, K^h_{m,y})$ is a reciprocal lattice vector of the
honeycomb lattice,
 $\mathbf{k}$ is a $\Lambda_{\theta,\beta}$-invariant point in
the Brillouin zone of the honeycomb lattice.
 We assume
$\mathbf{Q}_i$ is a $\Lambda_{\theta,\beta}$-invariant point in the
Brillouin zone. Substituting equation (\ref{trans}) and
$\mathbf{Q}_i=(Q_{ix}, Q_{iy})$ into equation  (\ref{invariant}), we
obtain  the following equation
\begin{eqnarray}
(Q_{ix},Q_{iy})=(-Q_{ix}-2\Delta_x(\mathbf{Q}_i),
-Q_{iy}+\pi-2\Delta_y(\mathbf{Q}_i))-(\cos\theta {K}^h_{m,x},
(1+\sin\theta)K^h_{m,y})
\end{eqnarray}
Solving the above equation, we obtain the
$\Lambda_{\theta,\beta}$-invariant points in the Brillouin zone are
$\mathbf{Q}_{1,2}=(\pm \arccos(-\beta/2)/\cos\theta,0)$. At the
$\Lambda_{\theta,\beta}$-invariant points $\mathbf{Q}_i (i=1,2)$, we
have
\begin{eqnarray}
\Psi'_{h,\mathbf{Q}_i}(\mathbf{r})=\Lambda_{\theta,\beta}\Psi_{h,\mathbf{Q}_i}(\mathbf{r})
\end{eqnarray}
It is easy to verify that
$\Lambda_{\theta,\beta}^2=\Omega_{\theta,\beta}^{-1}\circ\Upsilon^2\circ\Omega_{\theta,\beta}$.
Therefore, we have
$\Lambda_{\theta,\beta}^2\Psi_{h,\mathbf{k}}(\mathbf{r})=e^{-2i[k_x+\Delta_x(\mathbf{k})]}\Psi_{h,\mathbf{k}}(\mathbf{r})
$. At the $\Lambda_{\theta,\beta}$-invariant point $\mathbf{Q}_i$,
we have the following equation :
\begin{eqnarray}
({\Psi'_{h,\mathbf{Q}_i}},\Psi_{h,\mathbf{Q}_i})&=&(\Lambda_{\theta,\beta}\Psi_{h,\mathbf{Q}_i},\Lambda_{\theta,\beta}{\Psi'_{h,\mathbf{Q}_i}})
=(\Lambda_{\theta,\beta}\Psi_{h,\mathbf{Q}_i},\Lambda_{\theta,\beta}^2\Psi_{h,\mathbf{Q}_i})\nonumber\\
&=&e^{-2i[Q_{ix}+\Delta_x(\mathbf{Q}_i)]}( {\Psi'_{h,\mathbf{Q}_i}},
\Psi_{{h,\mathbf{Q}_i}}).
\end{eqnarray}
Substituting $\mathbf{Q}_{1,2}$, we obtain
$\Lambda_{\theta,\beta}^2=e^{-2i[Q_{ix}+\Delta_x(\mathbf{Q}_i)]}=-1$
at $\mathbf{Q}_{1,2}$. Thus, we have the solution
$({\Psi'_{h,\mathbf{Q}_i}},\Psi_{h,\mathbf{Q}_i}) =0$, which implies
that $\Psi_{h,\mathbf{Q}_i}$ and ${\Psi'_{h,\mathbf{Q}_i}}$ are
orthogonal to each other. We can conclude that there must be  the
band degeneracy at the points $\mathbf{Q}_{1,2}$.

\end{document}